\documentclass[11pt]{article}
\usepackage{moriond,epsfig}

\def\Journal#1#2#3#4{{#1} {\bf #2}, #3 (#4)}

\def\JPG{{\em J. Phys.} G}

\def\NPB{{\em Nucl. Phys.} B}
\def\EPJC{{\em Eur. Phys. J.} C}
\def\PLB{{\em Phys. Lett.}  B}

\def\PRD{{\em Phys. Rev.} D}

\def\be{\begin{equation}}
\def\ee{\end{equation}}
\def\bea{\begin{eqnarray}}
\def\eea{\end{eqnarray}}

\begin{document}
\vspace*{4cm}
\title{HEAVY FLAVOUR PRODUCTION AT HERA}

\author{ BENNO LIST\\For the H1 and ZEUS Collaborations }

\address{University of Hamburg, Institute for Experimental Physics,\\
Luruper Chaussee 149, D--22761 Hamburg}

\maketitle\abstracts{
The production of charm and beauty quarks in ep collisions at HERA
has been studied by the H1 and ZEUS collaborations.
Charm production is generally well described 
in total rate and in shape
by next to leading order
(NLO) 
calculations in perturbative quantum chromodynamics (QCD),
although in specific phase space corners
the NLO calculations underestimate the observed cross sections.
More and more
beauty production data are becoming available.
For this process, NLO
QCD predictions
tend to be lower than the measurements.
}

\section{Introduction}

The production of charm and beauty quarks
has been studied in great detail over the last years
at the ep collider HERA by the H1 and ZEUS collaborations.
In ep scattering, the main production mechanism 
of heavy quarks is boson gluon fusion,
where a pair of heavy quarks is formed in the collision
of a photon emitted by the electron\footnote{At different times,
HERA ran with electrons or positrons; here we 
use the term ``electron'' to denote either.} and a gluon
out of the proton.
The mass of the charm or beauty quark provides a hard
scale that makes calculations in perturbative QCD (pQCD) 
viable even in the absence of additional hard scales
that may be provided by the photon virtuality $Q^2$
or the transverse momentum $p_t$ of the quarks.

Calculations based on pQCD have generally been quite successful
in describing charm production data, while beauty production
rates tend to be underestimated by such calculations.
The challenges faced by theory include the treatment of
the quark mass in different kinematic regions,
the inclusion of the effects of the intrinsic gluon transverse momentum
$k_t$, and a quantitative description of heavy flavour production
in processes where the photon's hadronic structure is resolved.

\section{Charm Production}

A wealth of charm production measurements at HERA exists,
in both photoproduction, where the exchanged photon is quasi--real ($Q^2
\approx 0$), and deep-inelastic scattering (DIS), where the
photon virtuality $Q^2$ is large compared to $\Lambda_{QCD}$.
Mostly, these measurements are based on $D^*$ tagging,
where decays of the $D^{*+}$ meson\footnote{Charge conjugate states are
implicitly included in this text.} to $D^0 \pi^+$ with subsequent $D^0$ 
decays to 
$K^- \pi^+$ or $K^- \pi^+\pi^+\pi^-$ are used to identify 
charm production. 
This technique is well understood, but 
hampered by the rather small branching fractions of the decays involved.
Recently, also measurements that exploit lifetime information
using vertex detectors have been presented; these measurements profit 
from larger sample sizes and the possibility to measure charm and beauty 
production concurrently, but require a vertex detector,
whose resolution has to be well understood.

Calculations based on pQCD \cite{bib:ffns,bib:vfns}, using parton densities
of the proton derived from inclusive DIS measurements \cite{Chekanov:2002pv}, 
have been very successful in describing inclusive charm production
in photoproduction and DIS over many orders of magnitude in cross sections
\cite{bib:f2charm,Aktas:2004az,Aktas:2005iw}.
Of particular importance for these calculations is the gluon density,
which can be inferred from the scaling violations in the 
inclusive DIS measurements.
Still, at low values of $Q^2$,
the charm production data tends to be higher than the 
pQCD predictions. 
The experimental accuracy of these data is now sufficient
to further constrain the gluon density;
however, theoretical uncertainties in this region are still substantial
and thus
call for an improved theoretical understanding.

\begin{figure}[bthp]
\begin{center}
\setlength{\unitlength}{1cm}
\begin{picture}(13,7.5)
\put(0,0){\psfig{figure=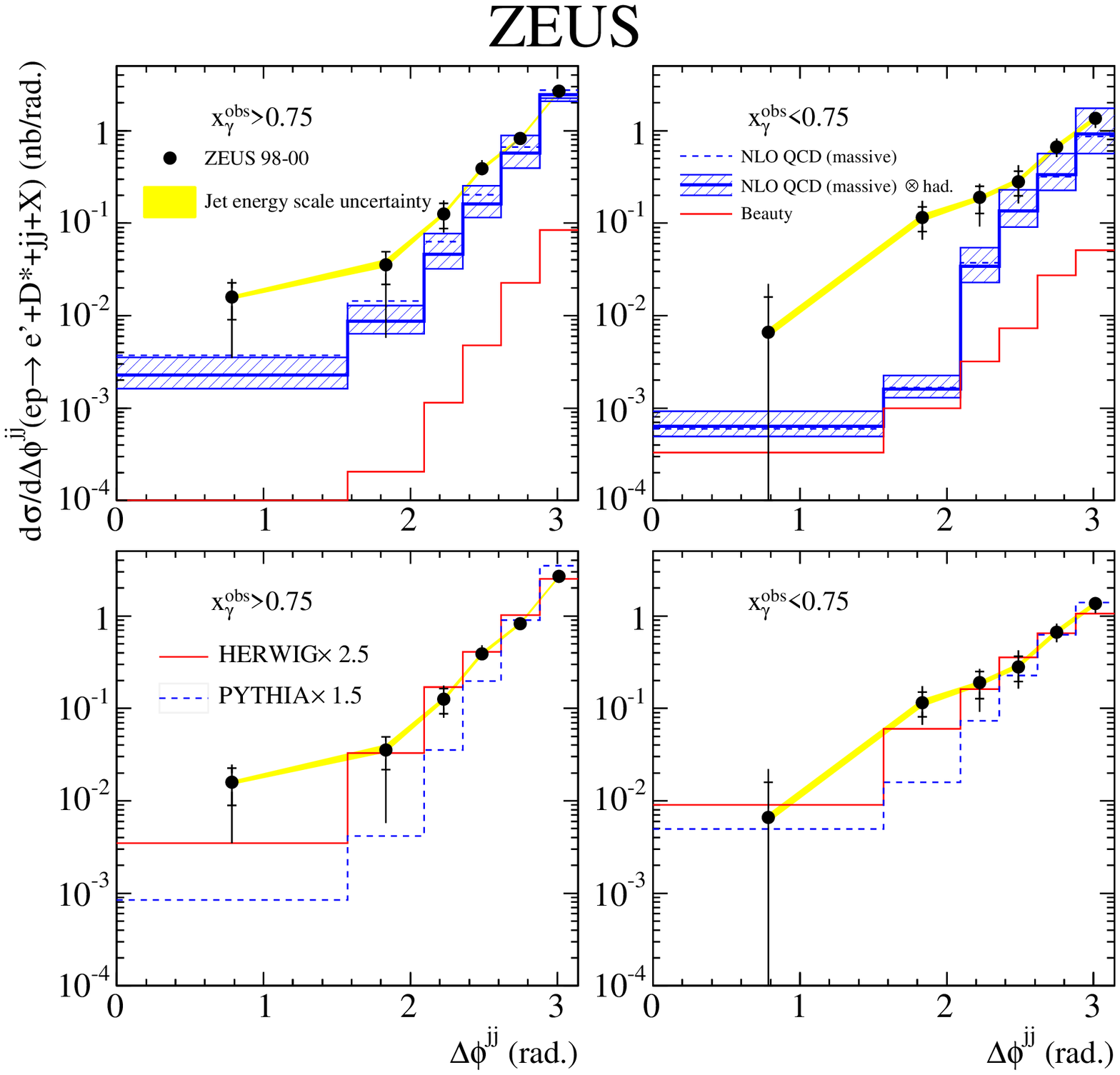,height=7.5cm}}
\put(8.5,3.75){\psfig{figure=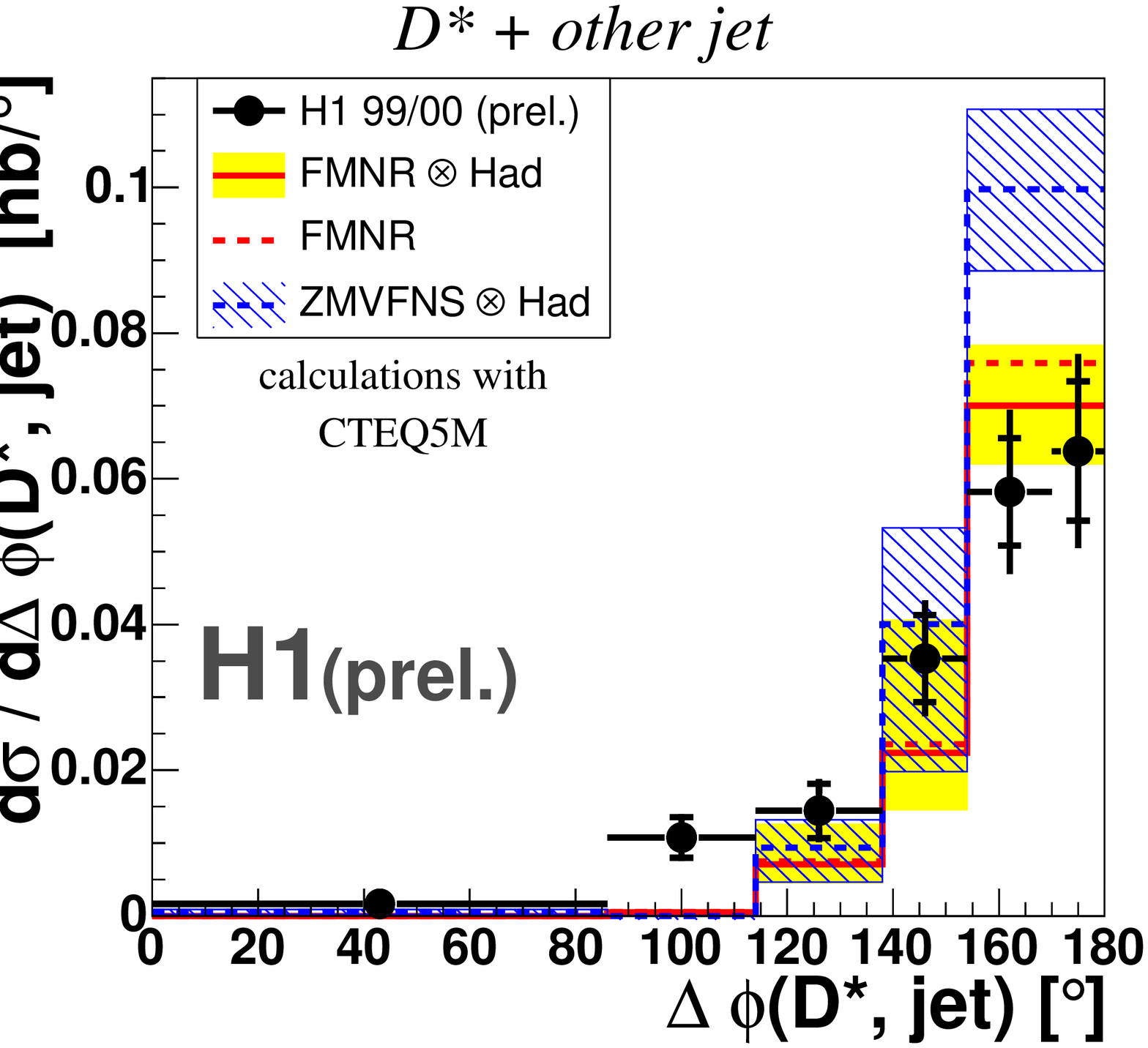,height=3.75cm}}
\put(8.5,0){\psfig{figure=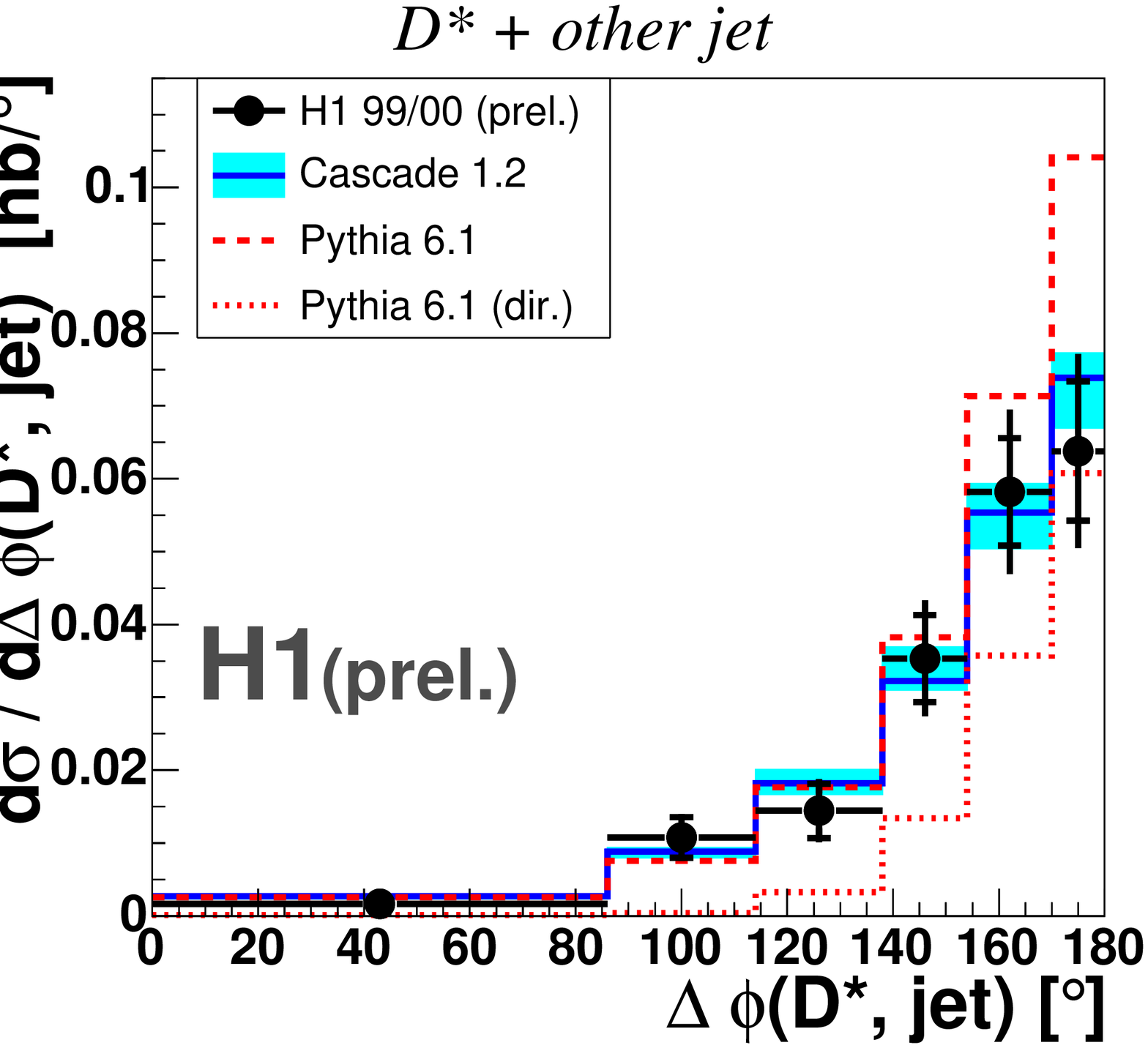,height=3.75cm}}
\end{picture}
\end{center}
\caption[]{ZEUS \protect\cite{Chekanov:2005zg} (left) and H1 \protect\cite{Flucke:2005ux} 
(right) measurements of 
charm production in events with a $D^*$ meson and jets. 
Shown is the opening angle
$\Delta \phi$ between the jets in the plane perpendicular to the beam,
in comparison to NLO QCD calculations (top) and 
Monte Carlo models (bottom).
}
\label{fig:dstarjets}
\end{figure}

Recently, a new $D^*$ measurement has been published by the ZEUS 
collaboration \cite{Chekanov:2005zg}
that 
focuses on charm photoproduction with jets,
and H1 has presented similar preliminary data \cite{Flucke:2005ux};
here, the jet $p_t$ provides a hard scale that dominates
over the charm quark mass. 
In addition, the jets lead to additional observables whose
distributions may be confronted with theoretical preditions.
Of particular interest here is
$x_\gamma$, which in a leading order picture 
corresponds to the fraction of the photon's momentum
that enters the hard interaction; 
the measured observable is defined
as $x_\gamma^{obs} = (E-p_z)_{jets}/(E-p_z)_{had}$,
i.e. the fraction of $E-p_z$ of the two leading jets
compared to the total $E-p_z$ of the hadronic final state.
This quantity is expected to be close to $1$ for
``direct'' processes, where the photon couples pointlike
to the heavy quark, while events where the photon's hadronic structure is
resolved tend to lie at lower values of $x_\gamma^{obs}$. 

The ZEUS collaboration has also measured the opening angle
$\Delta \phi^{jj}$ in the transverse plane, 
which is expected to be $\pi$ in leading order (order $\alpha_s$) QCD,
while radiation of additional gluons leads to smaller opening angles.
Thus, this measurement tests effects of order
$\alpha_s^2$, so that ``NLO'' calculations are in fact leading order 
for this observable. 
H1 has performed a similar measurement with events  
that contain a $D^*$ meson plus a separate jet,
where H1 has measured the opening angle between the $D^*$ and the 
jet directions. 
Both measurements are shown in Fig.~\ref{fig:dstarjets}.
While the region of large opening angles is well described
by the calculations, the NLO predictions fall below the data
when the jets are not completely back-to-back anymore,
in particular in the resolved regime ($x_\gamma^{obs}<0.75$), as the
ZEUS measurement shows. 
In contrast,
Monte Carlo models, which approximate higher orders by additional parton
showers, are more successful in describing these distributions.
A significant improvement is hoped for when Monte Carlo programs
will become available for ep scattering processes
that mach NLO calculations with
parton showers, such as MC@NLO \cite{bib:mcnlo}.

\section{Beauty Production}

The measurement of beauty production is considerably more difficult
compared to charm production due to the smaller 
cross section and the absence of easily identifiable 
decay channels with sizeable branching fractions.
Beauty production can be identified by the observation of
events with muons that have a large relative transverse momenta
$p_{t, rel}$ with respect to a nearby jet, or by the exploitation
of lifetime information; these techniques have also been combined.

Both, the ZEUS and H1 collaborations, have published results 
\cite{Chekanov:2004tk,Aktas:2005zc} for
visible beauty production cross sections
in events with muons and jets, which are characterized by
restrictions on the $p_t$ and pseudorapidity $\eta$ of the muons
and jets. 
Event with $D^*$ mesons and muons \cite{Aktas:2005bt}
or dimuons \cite{Longhin:2005vh}
have also be used to extract beauty cross sections.

\begin{figure}[bthp]
\begin{center}
\psfig{figure=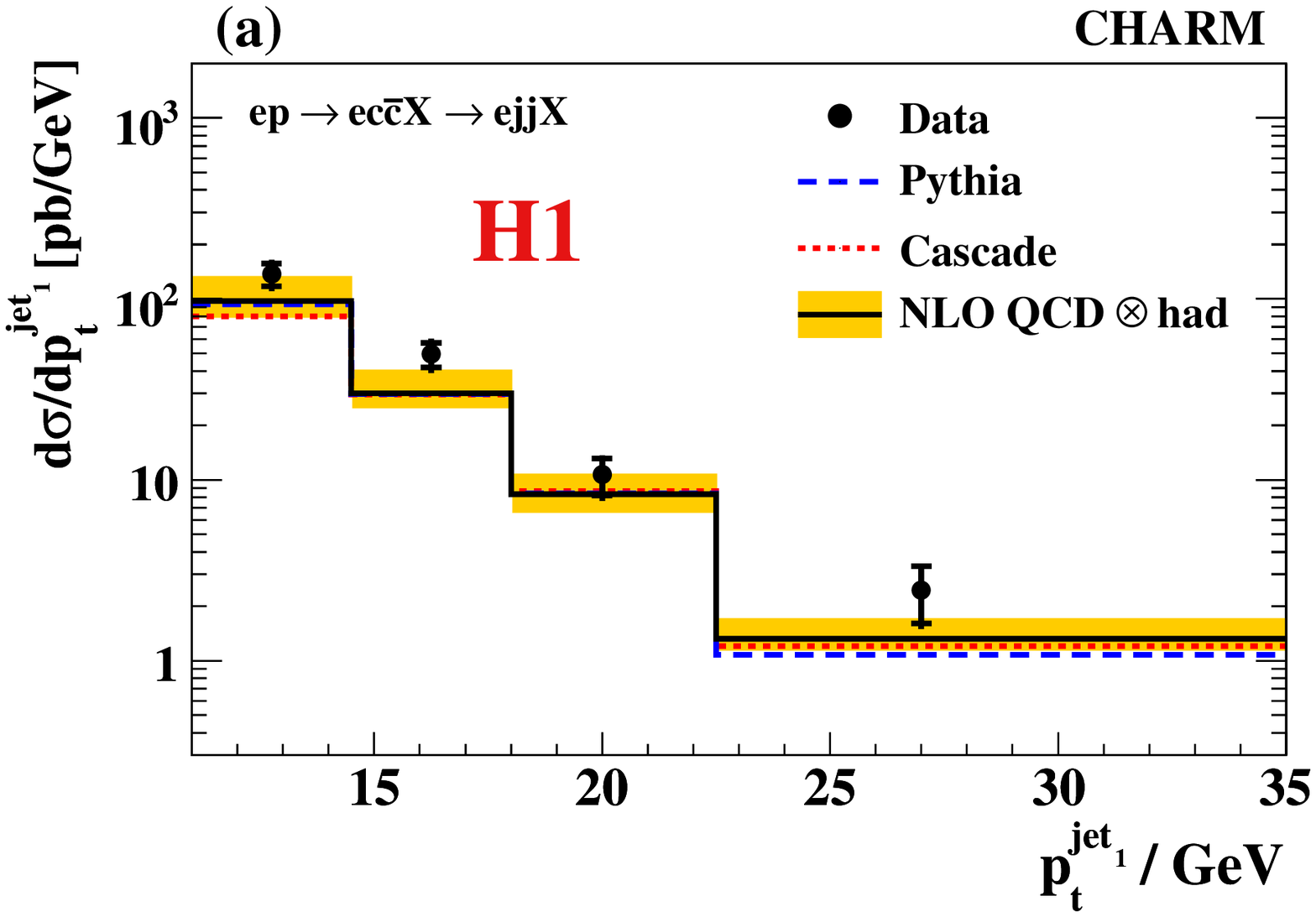,height=1.5in}
\psfig{figure=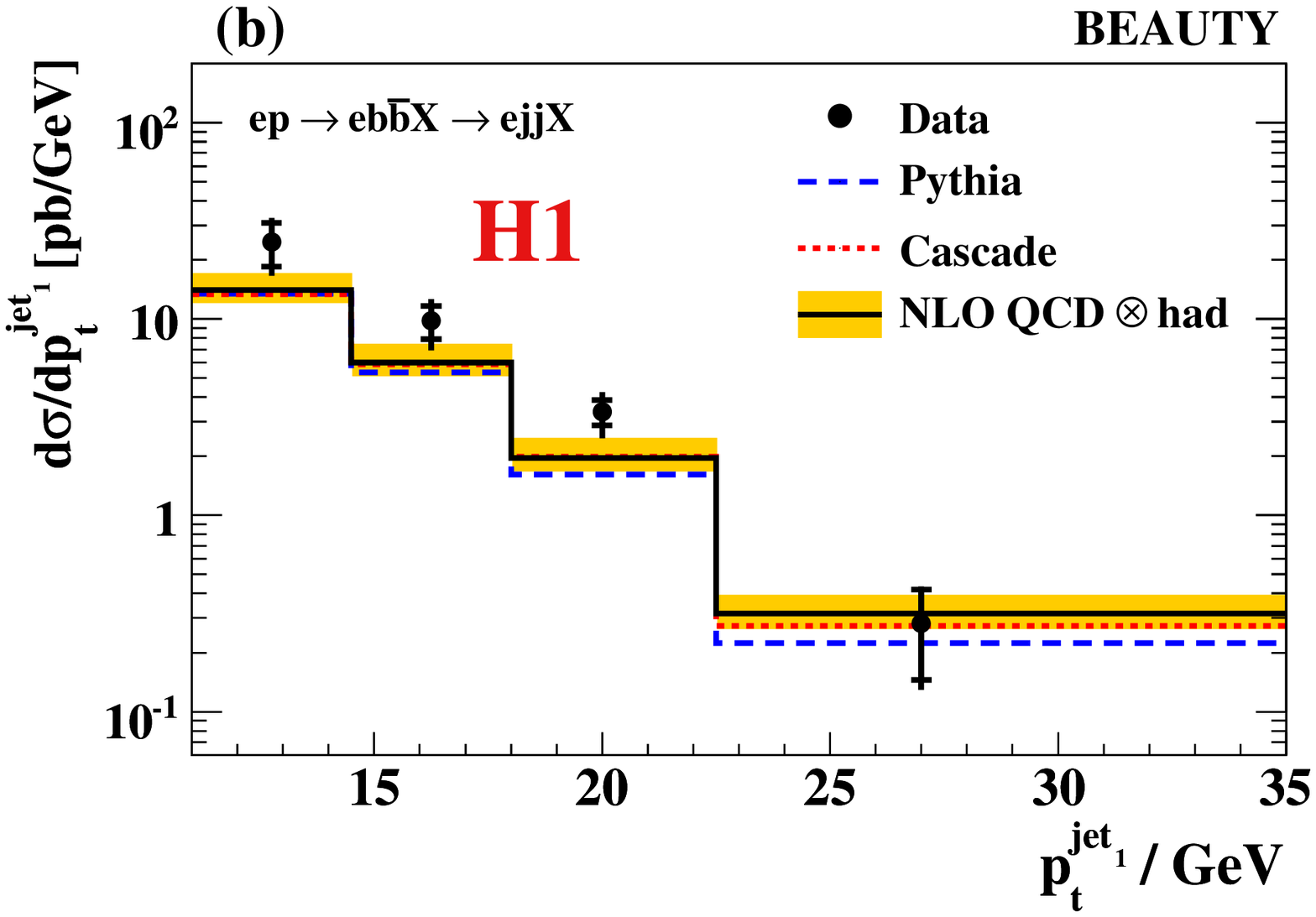,height=1.5in}\\
\vspace*{3mm}
\psfig{figure=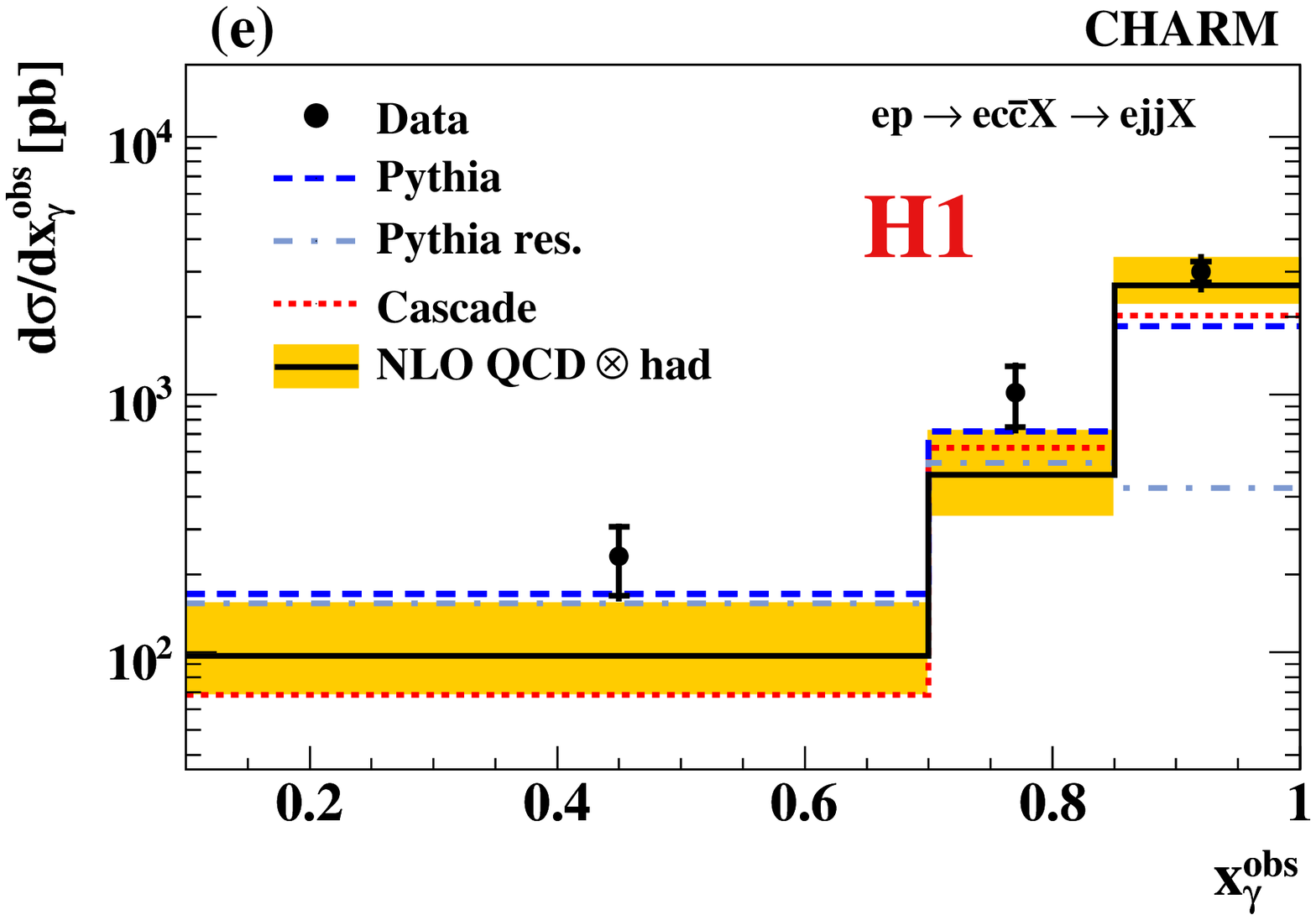,height=1.5in}
\psfig{figure=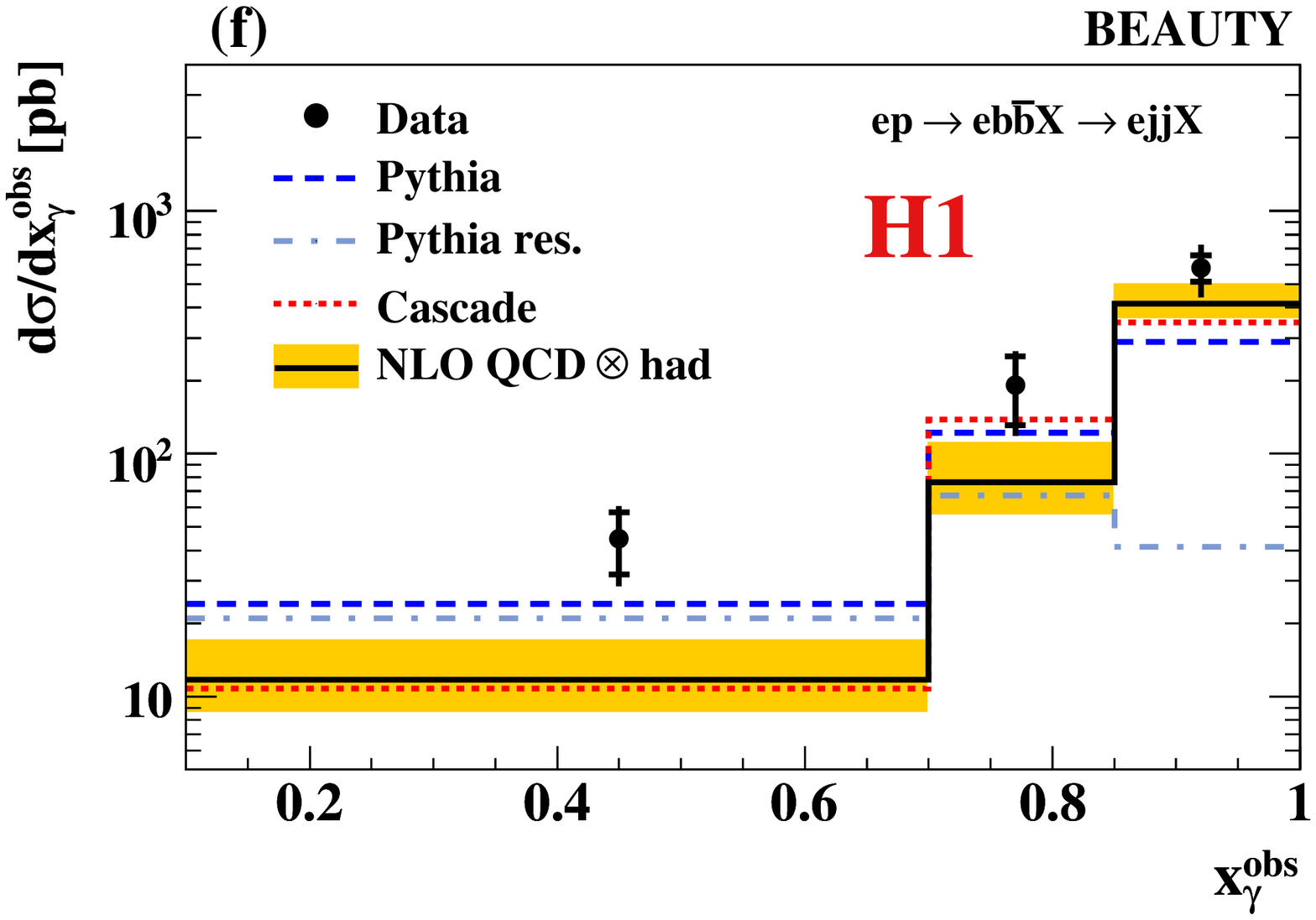,height=1.5in}
\end{center}
\caption[]{H1 measurements \protect\cite{bib:h1-bjets} of inclusive charm and 
beauty production
in photoproduction with two jets.}
\label{fig:h1-bjets}
\end{figure}

Recently, the H1 collaboration has also published inclusive 
beauty production cross sections
in DIS \cite{Aktas:2004az,Aktas:2005iw}, 
and in photoproduction with two jets
\cite{bib:h1-bjets}, 
based solely on lifetime
information; 
the ZEUS collaboration has also presented preliminary beauty 
production measurements based on their micro vertex detector
\cite{Hall-Wilton:2005vf}. 
This is also the first heavy flavour result from the HERA-II 
running phase, while the other results 
discussed in this presentation were achieved with HERA-I data.

The inclusive beauty production measurements
 in DIS are well discribed
by the NLO QCD calculations, and approach a precision
that is sufficient to discriminate between different parametrizations
of parton densities in the proton.
It is encouraging to see also NNLO calculations
of beauty production 
appear \cite{Thorne:2005nz}, which agree well with the 
DIS measurements.
On the other hand, the rate of beauty photoproduction
with two jets is significantly underestimated by the 
theory calculations, as shown in Fig.~\ref{fig:h1-bjets}.
The discrepancy is again particularly large in the region
of small $x_\gamma^{obs}<0.85$.

In general, a large fraction of the beauty production measurements
show higher cross sections than predicted by
QCD calculations; among observations where the discrepancy is
particularly significant are the 
H1 measurement of visible b cross sections with muons and jets in 
DIS and photoproduction \cite{Aktas:2005zc}, 
the new H1 dijet measurement in photoproduction \cite{bib:h1-bjets},
a measurement using $D^*\mu$ correlations \cite{Aktas:2005bt},
and a ZEUS measurement  of visible b cross sections with muons and jets 
in DIS \cite{Chekanov:2004tk}.
However, essentially all measurements are compatible
with cross section values that are approximately a factor 
$1.5$ to $1.7$ larger than the NLO QCD prediction, 
and several new measurements \cite{Aktas:2004az,Aktas:2005iw,Chekanov:2004xy}
are also consistent within errors
 with a ratio between data and theory of $1$.


\section*{References}
\begin{thebibliography}{99}
  
\bibitem{bib:ffns}
  B.~W.~Harris and J.~Smith,
  \Journal{\PRD}{57}{2806}{1998}; \\
  %
  E.~Laenen, S.~Riemersma, J.~Smith and W.~L.~van Neerven,
  \Journal{\NPB}{392}{162}{1993};
%
  \Journal{\em ibid.}{392}{229}{1993}.
  
\bibitem{bib:vfns}
  S.~Frixione, M.~L.~Mangano, P.~Nason and G.~Ridolfi,
  \Journal{\PLB}{348}{633}{1995}; \\
  %
  R.~S.~Thorne,
  \Journal{\JPG}{25}{1307}{1999}; \\
%
  R.~S.~Thorne and R.~G.~Roberts,
  \Journal{\EPJC}{19}{339}{2001}.

\bibitem{Chekanov:2002pv}
  S.~Chekanov {\it et al.}  [ZEUS Collaboration],
  \Journal{\PRD}{67}{012007}{2003}.

\bibitem{bib:f2charm}
  C.~Adloff {\it et al.}  [H1 Collaboration],
  \Journal{\PLB}{528}{199}{2002}; \\
%
  {\it idem},
  \Journal{\NPB}{545}{21}{1999}; \\
%
  J.~Breitweg {\it et al.}  [ZEUS Collaboration],
  \Journal{\PLB}{407}{402}{1997}; \\
%
  {\it idem},
  \Journal{\EPJC}{12}{35}{2000}; \\
%
  S.~Chekanov {\it et al.}  [ZEUS Collaboration],
  \Journal{\PRD}{69}{012004}{2004}; \\
%
  G.~Aghuzumtsyan  [ZEUS Collaboration],
  \Journal{\em AIP Conf.\ Proc.}{792}{803}{2005}.

\bibitem{Aktas:2004az}
  A.~Aktas {\it et al.}  [H1 Collaboration],
  \Journal{\EPJC}{40}{349}{2005}.

\bibitem{Aktas:2005iw}
  A.~Aktas {\it et al.}  [H1 Collaboration],
  \Journal{\EPJC}{45}{23}{2006}.

\bibitem{Chekanov:2005zg}
  S.~Chekanov {\it et al.}  [ZEUS Collaboration],
  \Journal{\NPB}{729}{492}{2005}.

\bibitem{Flucke:2005ux}
  G.~Flucke  [H1 Collaboration],
  \Journal{\em AIP Conf.\ Proc.}{792}{815}{2005}.

\bibitem{bib:mcnlo}  
  S.~Frixione and B.~R.~Webber,
  \Journal{\em JHEP}{0206}{029}{2002}; \\
%
  S.~Frixione, P.~Nason and B.~R.~Webber,
  \Journal{\em JHEP}{0308}{007}{2003}.

\bibitem{Chekanov:2004tk}
  S.~Chekanov {\it et al.}  [ZEUS Collaboration],
  \Journal{\PLB}{599}{173}{2004}.

\bibitem{Aktas:2005zc}
  A.~Aktas {\it et al.}  [H1 Collaboration],
  \Journal{\EPJC}{41}{453}{2005}.

\bibitem{Aktas:2005bt}
   A.~Aktas {\it et al.}  [H1 Collaboration],
  \Journal{\PLB}{621}{56}{2005}.

\bibitem{Longhin:2005vh}
  A.~Longhin  [ZEUS Collaboration],
  \Journal{\em AIP Conf.\ Proc.}{792}{887}{2005}.
  
\bibitem{bib:h1-bjets}
  A.~Aktas {\it et al.}  [H1 Collaboration],
  {\em arXiv:}hep-ex/0605016 (2006).

\bibitem{Hall-Wilton:2005vf}
  R.~J.~Hall-Wilton [ZEUS Collaboration],
  \Journal{\em AIP Conf.\ Proc.}{792}{879}{2005}.

\bibitem{Thorne:2005nz}
  R.~S.~Thorne,
  \Journal{\em AIP Conf.\ Proc.}{792}{847}{2005}.

\bibitem{Chekanov:2004xy}
  S.~Chekanov {\it et al.}  [ZEUS Collaboration],
  \Journal{\PRD}{70}{012008}{2004}.
  
\end{thebibliography}
\end{document}